\documentclass[twocolumn,letterpaper,aps,prl,superscriptaddress,showpacs,floatfix,longbibliography]{revtex4-1}

\usepackage{graphicx}	
\usepackage{xspace}	
\usepackage{multirow}

\newcommand{\pt}{\mbox{$p_T$}\xspace}

\newcommand{\dAu}{\mbox{$d$$+$Au}\xspace}
\newcommand{\pA}{\mbox{$p$+$A$}\xspace}
\newcommand{\pp}{\mbox{$p$+$p$}\xspace}
\newcommand{\pAu}{\mbox{$p$$+$Au}\xspace}
\newcommand{\pAl}{\mbox{$p$$+$Al}\xspace}
\newcommand{\polpp}{\mbox{$p^{\uparrow}$+$p$}\xspace}
\newcommand{\polpA}{\mbox{$p^{\uparrow}$+$A$}\xspace}
\newcommand{\polpAu}{\mbox{$p^{\uparrow}$+Au}\xspace}
\newcommand{\polpAl}{\mbox{$p^{\uparrow}$+Al}\xspace}

\begin{document}

\title{Nuclear dependence of transverse-single-spin asymmetries in the 
production of charged hadrons at forward rapidity in polarized $p$$+$$p$, 
$p$$+$Al, and $p$$+$Au collisions at $\sqrt{s_{_{NN}}}=200$ GeV}

\newcommand{\abilene}{Abilene Christian University, Abilene, Texas 79699, USA}
\newcommand{\augie}{Department of Physics, Augustana University, Sioux Falls, South Dakota 57197, USA}
\newcommand{\banaras}{Department of Physics, Banaras Hindu University, Varanasi 221005, India}
\newcommand{\barc}{Bhabha Atomic Research Centre, Bombay 400 085, India}
\newcommand{\baruch}{Baruch College, City University of New York, New York, New York, 10010 USA}
\newcommand{\bnlcoll}{Collider-Accelerator Department, Brookhaven National Laboratory, Upton, New York 11973-5000, USA}
\newcommand{\bnlphys}{Physics Department, Brookhaven National Laboratory, Upton, New York 11973-5000, USA}
\newcommand{\caucr}{University of California-Riverside, Riverside, California 92521, USA}
\newcommand{\charlesczech}{Charles University, Ovocn\'{y} trh 5, Praha 1, 116 36, Prague, Czech Republic}
\newcommand{\chonbuk}{Chonbuk National University, Jeonju, 561-756, Korea}
\newcommand{\cns}{Center for Nuclear Study, Graduate School of Science, University of Tokyo, 7-3-1 Hongo, Bunkyo, Tokyo 113-0033, Japan}
\newcommand{\colorado}{University of Colorado, Boulder, Colorado 80309, USA}
\newcommand{\columbia}{Columbia University, New York, New York 10027 and Nevis Laboratories, Irvington, New York 10533, USA}
\newcommand{\czechtech}{Czech Technical University, Zikova 4, 166 36 Prague 6, Czech Republic}
\newcommand{\debrecen}{Debrecen University, H-4010 Debrecen, Egyetem t{\'e}r 1, Hungary}
\newcommand{\elte}{ELTE, E{\"o}tv{\"o}s Lor{\'a}nd University, H-1117 Budapest, P{\'a}zm{\'a}ny P.~s.~1/A, Hungary}
\newcommand{\eszterhazy}{Eszterh\'azy K\'aroly University, K\'aroly R\'obert Campus, H-3200 Gy\"ongy\"os, M\'atrai \'ut 36, Hungary}
\newcommand{\ewha}{Ewha Womans University, Seoul 120-750, Korea}
\newcommand{\famu}{Florida A\&M University, Tallahassee, FL 32307, USA}
\newcommand{\fsu}{Florida State University, Tallahassee, Florida 32306, USA}
\newcommand{\gsu}{Georgia State University, Atlanta, Georgia 30303, USA}
\newcommand{\hiroshima}{Hiroshima University, Kagamiyama, Higashi-Hiroshima 739-8526, Japan}
\newcommand{\howard}{Department of Physics and Astronomy, Howard University, Washington, DC 20059, USA}
\newcommand{\ihepprot}{IHEP Protvino, State Research Center of Russian Federation, Institute for High Energy Physics, Protvino, 142281, Russia}
\newcommand{\illuiuc}{University of Illinois at Urbana-Champaign, Urbana, Illinois 61801, USA}
\newcommand{\inrras}{Institute for Nuclear Research of the Russian Academy of Sciences, prospekt 60-letiya Oktyabrya 7a, Moscow 117312, Russia}
\newcommand{\instpasczech}{Institute of Physics, Academy of Sciences of the Czech Republic, Na Slovance 2, 182 21 Prague 8, Czech Republic}
\newcommand{\isu}{Iowa State University, Ames, Iowa 50011, USA}
\newcommand{\jaea}{Advanced Science Research Center, Japan Atomic Energy Agency, 2-4 Shirakata Shirane, Tokai-mura, Naka-gun, Ibaraki-ken 319-1195, Japan}
\newcommand{\kek}{KEK, High Energy Accelerator Research Organization, Tsukuba, Ibaraki 305-0801, Japan}
\newcommand{\korea}{Korea University, Seoul, 02841}
\newcommand{\kurchatov}{National Research Center ``Kurchatov Institute", Moscow, 123098 Russia}
\newcommand{\kyoto}{Kyoto University, Kyoto 606-8502, Japan}
\newcommand{\lawllnl}{Lawrence Livermore National Laboratory, Livermore, California 94550, USA}
\newcommand{\losalamos}{Los Alamos National Laboratory, Los Alamos, New Mexico 87545, USA}
\newcommand{\lund}{Department of Physics, Lund University, Box 118, SE-221 00 Lund, Sweden}
\newcommand{\lyon}{IPNL, CNRS/IN2P3, Univ Lyon, Université Lyon 1, F-69622, Villeurbanne, France}
\newcommand{\maryland}{University of Maryland, College Park, Maryland 20742, USA}
\newcommand{\mass}{Department of Physics, University of Massachusetts, Amherst, Massachusetts 01003-9337, USA}
\newcommand{\michigan}{Department of Physics, University of Michigan, Ann Arbor, Michigan 48109-1040, USA}
\newcommand{\muhlenberg}{Muhlenberg College, Allentown, Pennsylvania 18104-5586, USA}
\newcommand{\nara}{Nara Women's University, Kita-uoya Nishi-machi Nara 630-8506, Japan}
\newcommand{\natmephi}{National Research Nuclear University, MEPhI, Moscow Engineering Physics Institute, Moscow, 115409, Russia}
\newcommand{\newmex}{University of New Mexico, Albuquerque, New Mexico 87131, USA}
\newcommand{\nmsu}{New Mexico State University, Las Cruces, New Mexico 88003, USA}
\newcommand{\northcg}{Physics and Astronomy Department, University of North Carolina at Greensboro, Greensboro, North Carolina 27412, USA}
\newcommand{\ohio}{Department of Physics and Astronomy, Ohio University, Athens, Ohio 45701, USA}
\newcommand{\ornl}{Oak Ridge National Laboratory, Oak Ridge, Tennessee 37831, USA}
\newcommand{\orsay}{IPN-Orsay, Univ.~Paris-Sud, CNRS/IN2P3, Universit\'e Paris-Saclay, BP1, F-91406, Orsay, France}
\newcommand{\peking}{Peking University, Beijing 100871, People's Republic of China}
\newcommand{\pnpi}{PNPI, Petersburg Nuclear Physics Institute, Gatchina, Leningrad region, 188300, Russia}
\newcommand{\riken}{RIKEN Nishina Center for Accelerator-Based Science, Wako, Saitama 351-0198, Japan}
\newcommand{\rikjrbrc}{RIKEN BNL Research Center, Brookhaven National Laboratory, Upton, New York 11973-5000, USA}
\newcommand{\rikkyo}{Physics Department, Rikkyo University, 3-34-1 Nishi-Ikebukuro, Toshima, Tokyo 171-8501, Japan}
\newcommand{\saispbstu}{Saint Petersburg State Polytechnic University, St.~Petersburg, 195251 Russia}
\newcommand{\seoulnat}{Department of Physics and Astronomy, Seoul National University, Seoul 151-742, Korea}
\newcommand{\stonybrkc}{Chemistry Department, Stony Brook University, SUNY, Stony Brook, New York 11794-3400, USA}
\newcommand{\stonycrkp}{Department of Physics and Astronomy, Stony Brook University, SUNY, Stony Brook, New York 11794-3800, USA}
\newcommand{\tenn}{University of Tennessee, Knoxville, Tennessee 37996, USA}
\newcommand{\titech}{Department of Physics, Tokyo Institute of Technology, Oh-okayama, Meguro, Tokyo 152-8551, Japan}
\newcommand{\tsukuba}{Tomonaga Center for the History of the Universe, University of Tsukuba, Tsukuba, Ibaraki 305, Japan}
\newcommand{\vandy}{Vanderbilt University, Nashville, Tennessee 37235, USA}
\newcommand{\weizmann}{Weizmann Institute, Rehovot 76100, Israel}
\newcommand{\wigner}{Institute for Particle and Nuclear Physics, Wigner Research Centre for Physics, Hungarian Academy of Sciences (Wigner RCP, RMKI) H-1525 Budapest 114, POBox 49, Budapest, Hungary}
\newcommand{\yonsei}{Yonsei University, IPAP, Seoul 120-749, Korea}
\newcommand{\zagreb}{Department of Physics, Faculty of Science, University of Zagreb, Bijeni\v{c}ka c.~32 HR-10002 Zagreb, Croatia}
\affiliation{\abilene}
\affiliation{\augie}
\affiliation{\banaras}
\affiliation{\barc}
\affiliation{\baruch}
\affiliation{\bnlcoll}
\affiliation{\bnlphys}
\affiliation{\caucr}
\affiliation{\charlesczech}
\affiliation{\chonbuk}
\affiliation{\cns}
\affiliation{\colorado}
\affiliation{\columbia}
\affiliation{\czechtech}
\affiliation{\debrecen}
\affiliation{\elte}
\affiliation{\eszterhazy}
\affiliation{\ewha}
\affiliation{\famu}
\affiliation{\fsu}
\affiliation{\gsu}
\affiliation{\hiroshima}
\affiliation{\howard}
\affiliation{\ihepprot}
\affiliation{\illuiuc}
\affiliation{\inrras}
\affiliation{\instpasczech}
\affiliation{\isu}
\affiliation{\jaea}
\affiliation{\kek}
\affiliation{\korea}
\affiliation{\kurchatov}
\affiliation{\kyoto}
\affiliation{\lawllnl}
\affiliation{\losalamos}
\affiliation{\lund}
\affiliation{\lyon}
\affiliation{\maryland}
\affiliation{\mass}
\affiliation{\michigan}
\affiliation{\muhlenberg}
\affiliation{\nara}
\affiliation{\natmephi}
\affiliation{\newmex}
\affiliation{\nmsu}
\affiliation{\northcg}
\affiliation{\ohio}
\affiliation{\ornl}
\affiliation{\orsay}
\affiliation{\peking}
\affiliation{\pnpi}
\affiliation{\riken}
\affiliation{\rikjrbrc}
\affiliation{\rikkyo}
\affiliation{\saispbstu}
\affiliation{\seoulnat}
\affiliation{\stonybrkc}
\affiliation{\stonycrkp}
\affiliation{\tenn}
\affiliation{\titech}
\affiliation{\tsukuba}
\affiliation{\vandy}
\affiliation{\weizmann}
\affiliation{\wigner}
\affiliation{\yonsei}
\affiliation{\zagreb}
\author{C.~Aidala} \affiliation{\michigan} 
\author{Y.~Akiba} \email[PHENIX Spokesperson: ]{akiba@rcf.rhic.bnl.gov} \affiliation{\riken} \affiliation{\rikjrbrc} 
\author{M.~Alfred} \affiliation{\howard} 
\author{V.~Andrieux} \affiliation{\michigan} 
\author{N.~Apadula} \affiliation{\isu} 
\author{H.~Asano} \affiliation{\kyoto} \affiliation{\riken} 
\author{B.~Azmoun} \affiliation{\bnlphys} 
\author{V.~Babintsev} \affiliation{\ihepprot} 
\author{N.S.~Bandara} \affiliation{\mass} 
\author{K.N.~Barish} \affiliation{\caucr} 
% ...
\collaboration{PHENIX Collaboration} \noaffiliation


\date{\today}


\begin{abstract}


We report on the nuclear dependence of transverse-single-spin asymmetries 
(TSSAs) in the production of positively charged hadrons in polarized 
$p^{\uparrow}+p$, $p^{\uparrow}+$Al and $p^{\uparrow}+$Au collisions at 
$\sqrt{s_{_{NN}}}=200$ GeV. The measurements have been performed at 
forward rapidity ($1.4<\eta<2.4$) over the range of transverse momentum 
($1.8<p_{T}<7.0$ GeV$/c$) and Feynman-$x$ ($0.1<x_{F}<0.2$). We observed 
positive asymmetries for positively charged hadrons in \polpp 
collisions, and significantly reduced asymmetries in $p^{\uparrow}$+$A$ 
collisions. These results reveal a nuclear dependence of TSSAs for 
charged-hadron production in a regime where perturbative techniques are 
applicable.  These results provide new opportunities to use \polpA 
collisions as a tool to investigate the rich phenomena behind TSSAs in 
hadronic collisions and to use TSSAs as a new handle in studying 
small-system collisions.

\end{abstract}

\maketitle


Understanding the transverse-single-spin asymmetries (TSSAs), that 
describe the azimuthal-angular dependence of particle production 
relative to the transverse-spin direction of the polarized proton in the 
reaction $p^{\uparrow}+p \rightarrow h + X$, has been a long-standing 
puzzle. The first observations in pion production at large Feynman-$x$ 
($x_F$)~\cite{klem:1976ui} showed measured TSSAs that were considerably 
larger than early theoretical predictions (in collinear leading twist 
approach)~\cite{Kane:1978nd}. Surprisingly large measured TSSAs 
continued to persist in hadronic collisions at high energies up to 
$\sqrt{s}=500~{\rm GeV}$~\cite{Antille:1980th,Adams:1991cs,Adams:1991rw,Allgower:2002qi,Adams:2003fx,Lee:2007zzh,Abelev:2008qb,Arsene:2008aa,Adamczyk:2012xd,Adare:2013ekj,Adare:2014qzo,Mondal:2014vla}. 
To explain these large TSSAs, two approaches were proposed within 
perturbative quantum chromodynamics (pQCD).  One approach is called 
transverse-momentum-dependent factorization, in which TSSAs are 
generated by correlations between the nucleon’s transverse spin 
direction and the transverse momentum of a parton in the polarized 
nucleon (Sivers effect~\cite{Sivers:1989cc,Sivers:1990fh}), and from the 
fragmentation of a transversely polarized parton into a final-state 
hadron (Collins effect~\cite{Collins:1992kk}). Another approach, 
directly applicable to single-hadron production (with 
$p_{T}\gg\Lambda_{\rm QCD}$) presented in this paper is the twist-3, 
collinear-factorization framework~\cite{Efremov:1981sh}. The full 
description of TSSAs in $\polpp\rightarrow h+X$ in the twist-3 collinear 
factorization includes twist-3 functions from the polarized proton, the 
unpolarized proton, and the parton fragmentation into final-state 
hadrons. The twist-3 functions describe quark-gluon-quark correlations 
and trigluon correlations in the polarized proton and have been studied in 
detail~\cite{Qiu:1998ia,Kouvaris:2006zy,Koike:2007rq,Koike:2009ge,Kanazawa:2010au,Kang:2011hk,Kanazawa:2011bg,Kang:2012xf,Beppu:2013uda}. 
Recently, calculations of the twist-3 contribution from parton 
fragmentation have been carried out and have shown this to be an 
important mechanism for understanding TSSA 
measurements~\cite{Metz:2012ct,Kanazawa:2014dca,Gamberg:2017gle}.

The Relativistic Heavy Ion Collider (RHIC) is a unique facility that can 
accelerate polarized protons and collide them with other (polarized) 
protons or nuclei~\cite{Alekseev:2003sk}. The extension of TSSA 
measurements to \polpA collisions not only gives us a crucial tool for 
understanding the nature of TSSAs, but also provides a new handle for 
studying \pA collisions and the parton dynamics inside nuclei, where 
many emergent effects remain to be understood. These include the 
so-called ``Cronin'' effect, an enhancement in the inclusive hadron \pt 
spectrum with respect to \pp collisions at moderate \pt of approximately 
$2<p_T<6$~GeV$/c$ that is proposed to be due to multiple scattering 
effects in the nuclear medium and modified hadronization
mechanism~\cite{Cronin:1974zm,Adler:2003ii,Aad:2016zif,Adare:2013esx}. 
Another exciting observation is that the collective behavior across 
large pseudorapidity ranges in high multiplicity \pA collisions that 
hints formation the quark-gluon 
plasma~~\cite{Dusling:2015gta,Nagle:2018nvi,PHENIX:2018lia}. 
Furthermore, when hadron production is measured in the proton-going 
direction, the properties of nuclear gluons in the small-$x$ region can 
be probed, where $x$ is the fraction of the proton's longitudinal 
momentum carried by the parton. The dynamics of gluons in the small-$x$ 
regime, where the gluon density is predicted to increase drastically, 
can be described by the color-glass condensate (CGC) 
formalism~\cite{Gelis:2010nm} at the saturation scale $Q_{s}$, where 
$Q^{2}_{sA}\propto A^{1/3}$ for the target 
nucleus~\cite{Schafer:2014zea, Zhou:2015ima}. In recent years, 
substantial attention has been given to an interplay between small-$x$ 
physics and spin physics by studying TSSAs in transversely-polarized 
proton and ion collisions (\polpA) and gluon saturation effects in a 
nucleus are taken into account for various calculations of TSSAs in 
\polpA collisions~\cite{Boer:2006rj, Kang:2011ni, Kovchegov:2012ga, 
Kang:2012vm, Zhou:2013gsa, Altinoluk:2014oxa, Schafer:2014zea, 
Zhou:2015ima, Boer:2015pni,Hatta:2016wjz,Hatta:2016khv,Benic:2018amn}. 
An $A$-dependence of TSSAs can arise from the $A$-dependence of 
$Q_{s}$ when the probe is at or below $Q_{s}$, while TSSAs are expected 
to be $A$-independent at higher 
scales~\cite{Boer:2006rj,Kang:2011ni,Hatta:2016wjz,Hatta:2016khv,Benic:2018amn}. 
Therefore, experimental data on hadron TSSAs measured in \pA collisions 
with varying $A$-size will provide valuable information testing these 
models and bring new insights in understanding the dynamics of the \pA 
collisions.

We report here on the observation of a nuclear dependence of TSSAs of 
positively-charged-hadron production at forward rapidity ($0.1<x_F<0.2$ 
and $1.4<\eta<2.4$, probing $0.004\lesssim x \lesssim 0.1$ in the 
nuclei) in collisions between transversely polarized protons and 
unpolarized protons or nuclei, \polpp, \polpAl, \polpAu at 
$\sqrt{s_{_{NN}}}=200{\rm\ GeV}$ measured with the PHENIX detector. The 
positively charged hadron is preferred in the nuclear-dependence 
measurement because the significance of TSSAs for negatively charged 
hadrons will be reduced by the partial cancellation of the asymmetry due 
to opposite signs of TSSAs for $\pi^{-}$ and $K^{-}$ in \polpp 
collisions~\cite{Arsene:2008aa,Lee:2007zzh}. In this measurement, we 
follow the convention to quantify TSSAs as $A_N$, where $A_N$ is the 
modulation of the azimuthal angle of the hadron ($\phi_h$) relative to 
the azimuthal angle of the transverse spin of the proton 
($\phi_{\rm pol}$), i.e., hadron-production cross section 
$\sigma \propto 1 + A_N \sin(\phi_{\rm pol} - \phi_h)$.


The data from transversely polarized \polpp, \polpAl, and \polpAu 
collisions at $\sqrt{s_{_{NN}}}=200\ $GeV were collected with the PHENIX 
detector during the RHIC 2015 running period. Proton beams were 
polarized vertically with respect to the beam direction with an average 
polarization of 58\% (clockwise-beam) or 57\% (counterclockwise-beam) 
for \polpp, 58\% for \polpAl, and 61\% for \polpAu collisions, with a 
relative uncertainty of 3\% due to uncertainty in the polarization 
normalization. The beams are bunched.  To minimize systematic effects 
due to time dependence of machine and detector performance, the spin 
configuration of the colliding bunches is alternated every 106 ns.

The PHENIX detector comprises two central arms at midrapidity and 
two muon arms at forward rapidity~\cite{Adcox:2003zm,Akikawa:2003zs}; only 
reconstructed tracks from the muon arms are used for this analysis. The 
two muon spectrometers cover $1.2<\eta<2.4$ (polarized $p$-going 
direction) and $-2.2<\eta<-1.2$ ($A$-going direction) in pseudorapidity 
with full azimuthal angle coverage. Each muon arm has 7.5 nuclear 
interaction lengths ($\lambda_{I}$) of hadron absorber followed by a 
muon tracker (MuTr), which is a set of three stations of cathode strip 
chambers for momentum measurements of charged particles. The MuTr 
determines the momentum of a charged particle in a radial magnetic field 
of $\int{B\cdot dl}=0.72~{\rm T\cdot m}$ with a momentum resolution of 
$\delta p/p\approx0.05$ for hadrons in the kinematic range of this 
analysis. A Muon Identifier (MuID), located behind the MuTr, comprises
five layers of stainless-steel absorbers ($\sim5\lambda_{I}$ total) 
and Iarocci tube planes. The MuID helps to identify muons and hadrons 
based on the penetration depth of the tracks at $p_z \gtrsim 3.5$ 
GeV$/c$~\cite{Adare:2012px}.

The beam-beam counters (BBCs)~\cite{Allen:2003zt}, at $z=\pm144~{\rm 
cm}$ from the nominal interaction point, comprise two arrays of 64 
quartz Cherenkov detectors and cover the full azimuth and the 
pseudorapidity range $3.1<|\eta|<3.9$. The BBCs are used to determine 
the collision vertex $z$-position ($|z|<30~{\rm cm}$ cut was used in 
this analysis) as well as to provide a minimum-bias (MB) trigger with 
efficiencies of 55\% for \pp, 72\% for \pAl, and 84\% for \pAu 
collisions. The $A$-going side of the BBC is also used to determine the 
event centrality based on the distribution of the charge 
sum~\cite{Adare:2013nff}. The recorded events are sampled by the MB 
trigger combined with muon triggers to enrich good muon and hadron 
tracks. The MuID provides a trigger for events containing one or more 
hadron or muon candidates. Momentum-sensitive triggering is provided by 
hit information from the MuTr to enrich tracks with $\pt>3~{\rm 
GeV}/c$~\cite{Adachi:2013qha}.


This analysis uses only charged tracks that stop in the middle of the 
MuID planes (third or fourth plane out of five planes) due to a hadronic 
interaction with the absorber material.  In the kinematic region 
of $0.1<x_F<0.2$, where the longitudinal momentum of particles is larger 
than $10~{\rm GeV}/c$, positively-charged hadron candidates mostly 
comprise $\pi^{+}$ and $K^{+}$.

The particle composition in the measured charged-hadron sample was 
estimated with a method developed in 
Ref.~\cite{Adare:2012px,Adare:2013lkk}, based on identified 
charged-hadron spectra measured at midrapidity in \pp and \dAu 
collisions at RHIC~\cite{Adare:2011vy,Agakishiev:2011dc,Adare:2013esx}, 
and extrapolated to PHENIX muon arm rapidity region of $1.2<\eta<2.4$ 
for \pp, \pAl and \pAu collisions using {\sc 
pythia}~\cite{Sjostrand:2006za} and {\sc hijing}~\cite{Gyulassy:1994ew} 
event generators. The $K^{+}/\pi^{+}$ ratio of $\sim0.35$, as measured 
at RHIC at midrapidity at $\pt\sim2~{\rm GeV}/c$ (typical for our 
data)~\cite{Adare:2011vy,Agakishiev:2011dc,Adare:2013esx}, was found 
approximately unchanged when extrapolated to forward rapidity in both 
\pp and \pA collisions. The $p/\pi^{+}$ ratio of $\sim0.25$ ($\sim0.35$) 
at midrapidity in \pp (\dAu) 
collisions~\cite{Adare:2011vy,Agakishiev:2011dc,Adare:2013esx} was 
extrapolated to the value of $\sim0.3$ ($\sim0.5$) at the muon arm 
rapidity, with ratios in \pAl and \pAu collisions being in between 
values for \pp and \dAu collisions. The initial charged hadron 
composition is significantly modified due to particle interaction in the 
detector material, which according to {\sc 
geant4}-based~\cite{Allison:2016lfl} detector simulation modifies the 
initial $K^{+}/\pi^{+}$ ($p/\pi^{+}$) ratio by a factor of 2.7 (0.4), 
which varies by ${\approx}5\%$ for different hadron interaction 
models~\cite{Allison:2016lfl}. As a result, the $\pi^{+}$/$K^{+}$/$p$ 
particle composition in our measured positively charged-hadron sample is 
evaluated to be 45\%/47\%/5\% in \pp collisions, with increased proton 
fraction to 7\% (9\%) in \pAl (\pAu) collisions.

The unbinned maximum-likelihood method for extracting $A_N$ was 
established in a previous study~\cite{Aidala:2017pum} which used the 
same detectors. Compared to binned approaches, this method is robust 
even for low-statistics data.  
The extended log-likelihood is defined to be
\begin{equation}
\label{eq:maxlikelihood}
\log \mathcal{L} = \sum_i \log(1 + P \cdot A_N \sin(\phi_{\rm pol} - \phi_h^i)) + \textrm{constant},
\end{equation}
where $\phi_h^i$ is the azimuthal angle of the $i$-th hadron with 
respect to the direction of the polarized proton beam, $\phi_{\rm pol}$ 
is the azimuthal angle for the beam polarization direction, which in the 
2015 PHENIX run takes the values $+/-\frac{\pi}{2}$ for 
$\uparrow/\downarrow$ spin-signed beam bunches, respectively, and $P$ is 
the beam polarization. The asymmetry $A_N$ is determined by maximizing 
$\log \mathcal{L}$. For \polpp collisions, both beams are polarized, 
therefore the values of $A_N$ were measured separately for each beam,
found to be consistent, and were averaged in the final result.  
For \polpA collisions, only the clockwise proton beam was 
polarized.  The statistical uncertainty was calculated from the second 
derivative of the log-likelihood estimator,

\begin{equation}
\label{eq:maxlikelihood_err}
\sigma^2(A_N) = (-\frac{\partial^2 \mathcal{L}}{ (\partial A_N)^2})^{-1}.
\end{equation}

The $A_N$ calculated from the likelihood method is compared with the 
following azimuthal-fitting method based on the polarization 
formula~\cite{Ohlsen:1973wf}:
\begin{equation}
\label{eq:anvsphi_pp}
A_N(\phi)=\frac{\sigma^{\uparrow}(\phi)-\sigma^{\downarrow }(\phi)}
{\sigma^{\uparrow}(\phi)+\sigma^{\downarrow}(\phi)}\\
=\frac{1}{P}\cdot\frac{N^{\uparrow}(\phi)
-R\cdot N^{\downarrow}(\phi)}{N^{\uparrow}(\phi)
+R\cdot N^{\downarrow}(\phi)},\\
\end{equation}
where $A_N(\phi)$ is the simple count-based transverse single-spin 
asymmetry in each of the 16 azimuthal $\phi$-bins, $\sigma^{\uparrow }$, 
$\sigma^{\downarrow}$ are cross sections for each polarization of spin 
up or down, $N^{\uparrow}$, $N^{\downarrow}$ are yields, and $R = 
L^{\uparrow}/L^{\downarrow}$ is the luminosity ratio (relative 
luminosity) between bunches with spin up and down, determined from the 
number of sampled MB triggers corresponding to different spin 
orientations. From this, $A_N$ is extracted from the fit of Eq. (3) 
with a function $A_N\cdot \sin(\phi_{\rm pol\uparrow} - \phi)$, where 
$\phi_{\rm pol \uparrow}=\pi/2$ is the azimuthal direction of the upward 
polarized bunches. Because every detector element is simultaneously used 
for the measurements with spin up and down, the possible systematic 
effects from acceptance nonuniformity and acceptance variation versus 
time are largely canceled. The relative variation between this 
method and the log likelihood method is included in the systematic 
uncertainty.

\begin{figure}[thb]
\includegraphics[width=0.995\linewidth]{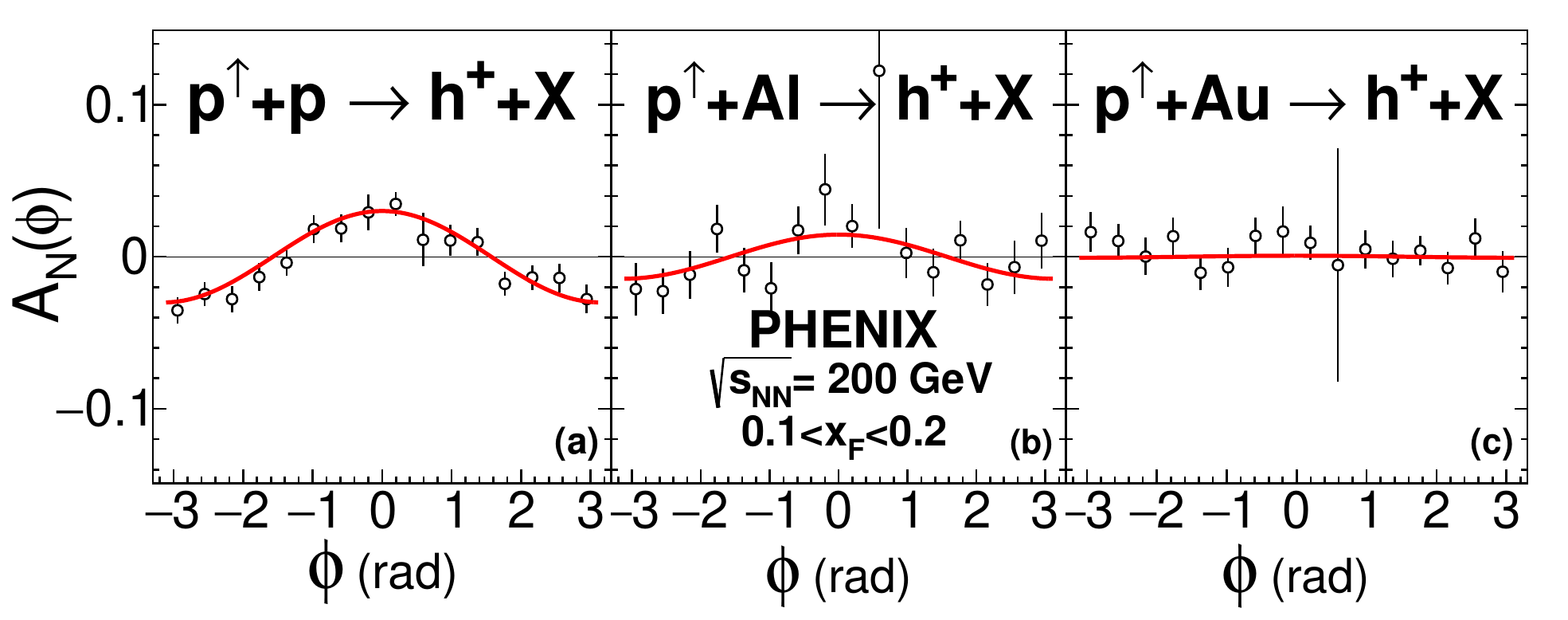}
\caption{
Azimuthal modulation of positively-charged hadrons for $1.4<\eta<2.4$, 
$0.1<x_F<0.2$, and $1.8<p_{T}<7.0$ GeV$/c$ in 
(a) \polpp, (b) \polpAl, and (c) \polpAu collisions 
at $\sqrt{s_{_{NN}}}=200~{\rm\ GeV}$.
}
\label{fig:plot_modulation}
\end{figure}

Figure~\ref{fig:plot_modulation} shows the reconstructed azimuthal 
modulation of positively-charged hadrons for $0.1<x_F<0.2$ and 
$1.8<p_T<7.0 \rm{\ GeV/}c$ in \polpp, \polpAl, and \polpAu collisions at 
$\sqrt{s_{_{NN}}}=200~{\rm GeV}$, as calculated using 
Eq.~(\ref{eq:anvsphi_pp}). The relatively larger statistical uncertainty 
in the bin at $\phi\sim0.6$~rad is caused by a known detector 
inefficiency. The $\chi^{2}/NDF$ of the fits are 10.1/15 for \polpp, 
13.5/15 for \polpAl, and 9.8/15 for \polpAu. The \polpp results show a 
clear nonzero modulation, while the \polpAl results show a weaker 
modulation. In \polpAu collisions, the modulation is consistent with 
zero within the statistical uncertainty.

The finite momentum and azimuthal angle $\phi$ resolution in the MuTr 
and the interactions of particles with the materials prior to entering 
the MuTr lead to a kinematic smearing for the $A_N$ measurement. This 
smearing effect was studied and corrected with a full detector {\sc 
geant4} simulation. The effect due to the $\phi$ smearing was found to 
be negligible. The momentum smearing effect was evaluated by resolving a 
set of linear equations connecting $A_N$ for the true $x_F$ bins 
($A_N^{\rm truth}$) and $A_N$ for the reconstructed $x_F$ bins 
($A_N^{\rm reco}$):
\begin{equation}
\label{eq:unfold1}
A_N^{{\rm reco}, m} =  \sum_{i} f^{i \rightarrow m} \cdot A_N^{{\rm truth}, i},
\end{equation}
where $A_N^{{\rm reco}, m}$ is $A_N$ for the $m$-th reconstructed $x_F$ 
bin from this measurement and $A_N^{{\rm truth}, i}$ is that for the 
$i$-th true $x_F$ bin. $f^{i \rightarrow m}$ represents the fraction of 
charged particles whose true $x_F$ at the collision vertex belongs to 
the $i$-th true $x_F$ bin and is reconstructed as being in the $m$-th 
$x_F$ bin. $f^{i \rightarrow m}$ is obtained from the {\sc geant4} 
detector simulation.  For calculating $A_N^{\rm truth}$ by solving 
Eq.~(\ref{eq:unfold1}), the $A_N^{\rm reco}$ is measured in a wider $x_F$ 
range $0.035<x_F<0.3$, by including two bins at lower $x_F$ and one bin 
at higher $x_F$. The resulting smearing-corrected $A_N^{\rm truth}$ of 
the positively-charged hadrons in bin $0.1<x_F<0.2$ are shown in 
Table~\ref{table:AN}. The difference between the obtained $A_N^{\rm 
truth}$ and the measured $A_N^{\rm reco}$ is small compared to the 
statistical uncertainty and is accounted for in the systematic 
uncertainty.

\begin{table}
\caption{\label{table:AN}
$A_N$ and sources of systematic uncertainty 
for positively-charged hadrons for $1.4<\eta<2.4$, $0.1<x_F<0.2$, 
and $1.8<p_{T}<7.0$ GeV$/c$ in \polpp, \polpAl, and \polpAu collisions 
at $\sqrt{s_{_{NN}}}=200~{\rm GeV}$.}
\begin{ruledtabular} \begin{tabular}{cccc}
  &   \polpp          &  \polpAl            &   \polpAu \\ \hline
$A_N$                   & $3.14\times 10^{-2}$ & $1.42\times 10^{-2}$ & $0.12\times 10^{-2}$ \\
$\delta A_N^{\rm stat}$ & $0.37\times 10^{-2}$ & $0.72\times 10^{-2}$ & $0.55\times 10^{-2}$ \\
$\delta A_N^{\rm syst}$ & $^{+0.05}_{-0.18}\times 10^{-2}$ & $^{+0.02}_{-0.02}\times 10^{-2}$ & $^{+0.06}_{-0.06}\times 10^{-2}$ \\
$\delta A_N^{\rm method}$   & $^{+0.05}_{-0.05}\times 10^{-2}$ & $^{+0.02}_{-0.02}\times 10^{-2}$ & $^{+0.06}_{-0.06}\times 10^{-2}$ \\
$\delta A_N^{\rm smear}$    & $^{+0.00}_{-0.17}\times 10^{-2}$ & $^{+0.01}_{-0.00}\times 10^{-2}$ & $^{+0.01}_{-0.00}\times 10^{-2}$ \\
\end{tabular} \end{ruledtabular}
\end{table}

\begin{figure*}[thb]
\includegraphics[width=0.45\linewidth]{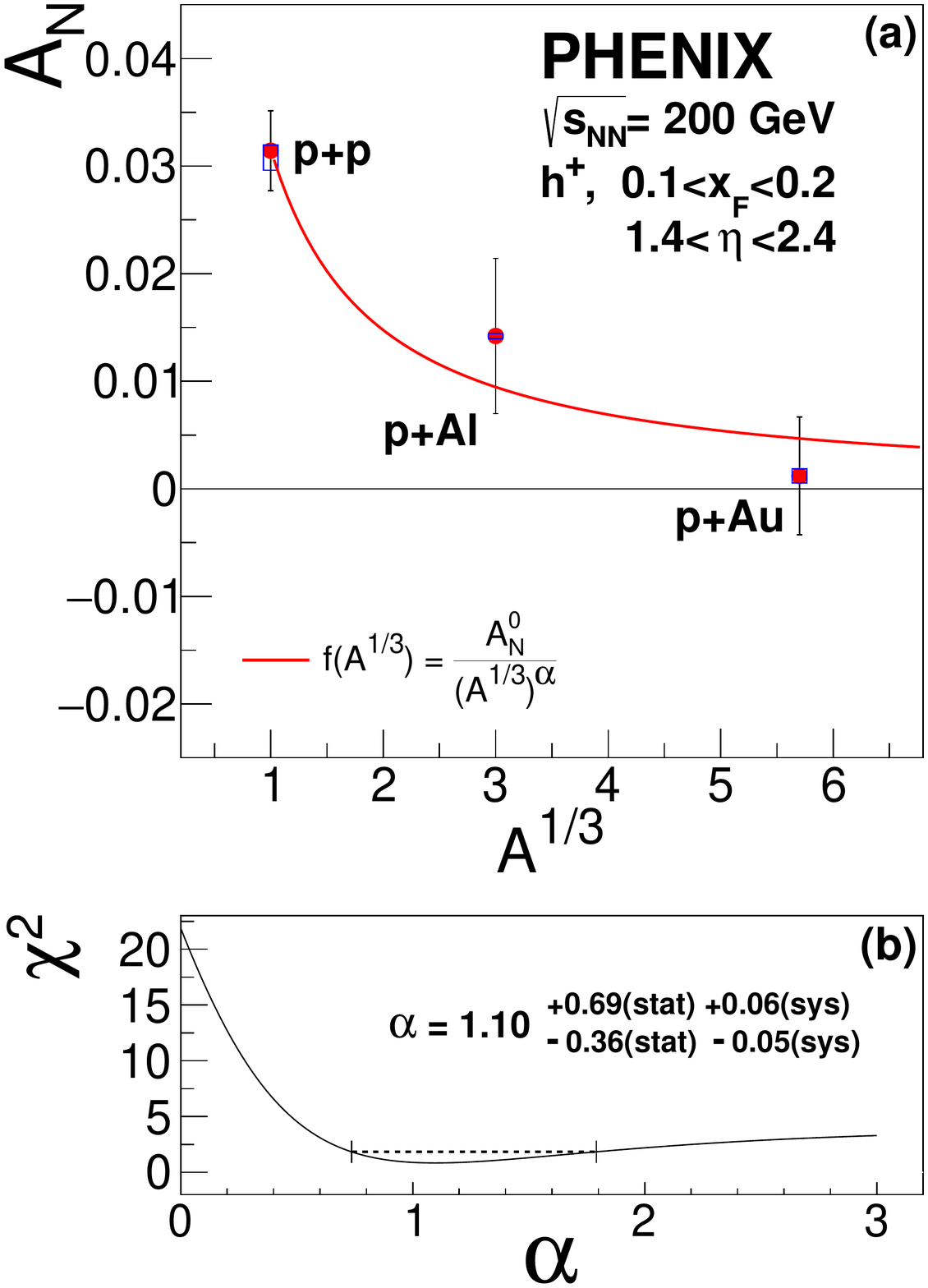}
\includegraphics[width=0.45\linewidth]{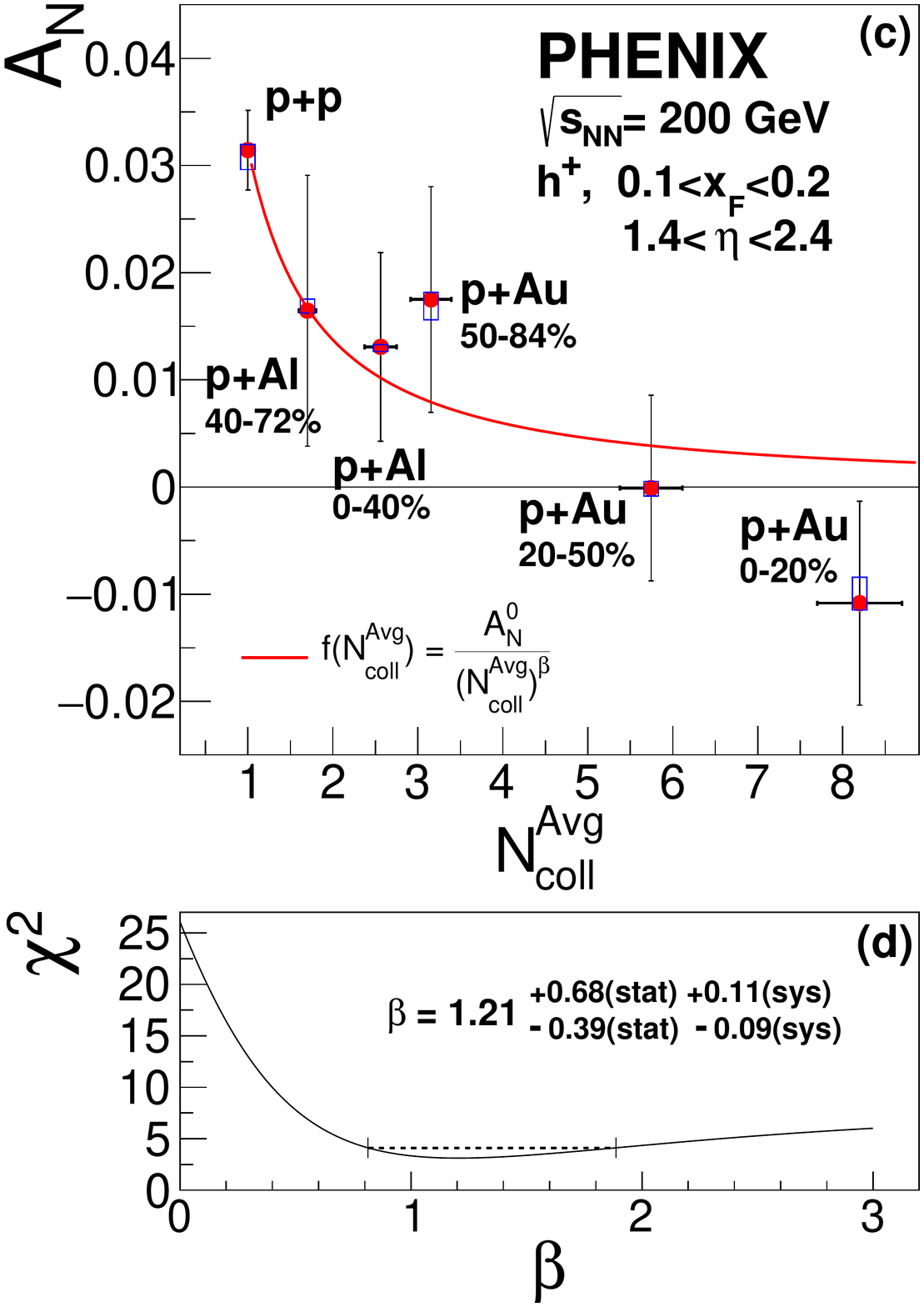}
\caption{
Upper panels are $A_N$ of positively-charged hadrons for $0.1<x_F<0.2$, 
$1.8<p_{T}<7.0$ GeV$/c$, and $1.4<\eta<2.4$ in \polpp, \polpAl, and 
\polpAu collisions at $\sqrt{s_{_{NN}}}=200~{\rm GeV}$ as a function of 
(a) $A^{1/3}$ and (c) $N_{\rm coll}^{\rm avg}$.  The fit functions, 
$A_N^{0}/(A^{1/3})^{\alpha}$ and $A_N^{0}/(N_{\rm coll}^{\rm 
avg})^{\beta}$ are shown as solid [red] curves.  Vertical bars (boxes) 
represent statistical (systematic) uncertainties. A 3\% scale 
uncertainty due to polarization uncertainty is not shown.  Lower panels 
show $\chi^{2}$ distributions as a function of power parameters (b) 
$\alpha$ and (d) $\beta$, taking into account the statistical 
uncertainty only. Dashed lines represent the range of $\alpha$ and 
$\beta$ for $\Delta\chi^{2}<1$.
}
\label{fig:plot_Adep}
\end{figure*}

Table~\ref{table:AN} also summarizes the systematic uncertainties for 
the $A_{N}$ measurements. The difference of $A_N$ extracted with two 
methods, Eqs.~(\ref{eq:maxlikelihood}) and~(\ref{eq:anvsphi_pp}), is 
shown as $\delta A_N^{\rm method}$. The difference between the obtained 
$A_N^{\rm truth}$ and measured $A_N^{\rm reco}$ is assigned as a 
conservative systematic uncertainty due to the smearing effect, 
$\delta A_N^{\rm smear}$. The total systematic uncertainty 
$\delta A_N^{\rm syst}$ is calculated as a quadratic sum of these two 
uncertainties.


Figure~\ref{fig:plot_Adep} shows $A_N$ of positively-charged hadrons in 
\polpp, \polpAl, and \polpAu collisions vs $A^{1/3}$ and the average 
number of nucleon-nucleon collisions $N_{\rm coll}^{\rm avg}$.  The 
$N_{\rm coll}^{\rm avg}$ is calculated using the Glauber 
model~\cite{Miller:2007ri} for each centrality class in \polpA 
collisions~\cite{Adare:2013nff}. The figure caption and legends denote 
the ranges of parameters and give the determined values of the power 
parameters $\alpha$ and $\beta$. Panels (b) and (d) show the $\chi^{2}$ 
distributions with only statistical uncertainties included.

The recent efforts to calculate $A_N$ in \polpp and \polpA collisions, 
accounting for gluon saturation effects 
~\cite{Gamberg:2017gle,Hatta:2016wjz,Hatta:2016khv,Benic:2018amn} 
suggested that $A_N$ could be $A$-independent or $A^{-1/3}$-dependent 
for the different contributions to $A_N$ in the region where $p_T<Q_s$. 
However, $\langle p_T\rangle\sim 2.9~{\rm GeV}/c$ in our results is much 
larger than the saturation scale in the Au nucleus ($Q_s^{\rm Au}\sim 0.9$ 
GeV) for the kinematics of this measurement and would lead to no 
strong $A$ dependence of TSSAs under these models, as calculated 
in Ref.~\cite{Benic:2018amn}.  Nevertheless, the results in this paper 
strongly disfavor the $A$-independent scenario.

The $N_{\rm coll}^{\rm avg}$-dependence of $A_N$ also suggests the 
decrease of $A_N$ is related to the density of nuclear matter inside the 
target nucleus which the projectile proton traverses. This 
$N_{\rm coll}^{\rm avg}$-dependence of $A_N$ could be related to novel 
effects in \pA collisions, such as multiple scattering of partons in the 
initial and/or final stages of the hard scattering, which is also 
indicated in the recent results of the nuclear modification of single 
hadron production and transverse momentum broadening in dihadron 
correlations in \pA 
collisions~\cite{Adare:2013esx,Aidala:2018eqn,Aidala:2019cnn}. 
Another possibility is interaction of the parton with hot QCD matter 
produced in \pA collisions, as suggested by recent results in small 
systems~\cite{Dusling:2015gta,Nagle:2018nvi,PHENIX:2018lia}.

We note preliminary results from the STAR 
collaboration~\cite{Heppelmann:2016siw} of measured $A_N$ for $\pi^0$ in 
\polpp and \polpAu collisions in more forward kinematics at 
$2.6<\eta<4.0$, $0.2<x_F<0.7$, and $p_T>1.5$~GeV/$c$ that show small or 
no $A$-dependence. The dramatic difference in $A$-dependence of TSSAs in 
different particle species and kinematic range emphasizes the importance 
of further detailed studies of $A_N$ for different particle species over 
wide kinematics.


To summarize, we have reported $A_N$ of positively-charged hadrons for 
$1.4<\eta<2.4$, $0.1<x_F<0.2$, and $1.8<p_{T}<7.0$ GeV$/c$ in \polpp, 
\polpAl, and \polpAu collisions at $\sqrt{s_{_{NN}}} = 200\ $GeV. For 
the first time, we observed an $A$-dependent $A_N$ in light hadron 
production in $p$$+$$A$ collisions, with the asymmetry values dropping 
from $\sim$3\% in \pp collisions to a value consistent with zero in 
$p$$+$Au collisions. These results may provide new insights into the 
origin of $A_N$ and a unique tool to investigate the rich phenomena 
behind TSSAs in hadronic collisions and to use TSSAs as a new approach to 
studying the small-system collisions.




\begin{acknowledgments}

We thank the staff of the Collider-Accelerator and Physics
Departments at Brookhaven National Laboratory and the staff of
the other PHENIX participating institutions for their vital
contributions.  
We also thank D. Pitonyak and M. Sievert for very useful discussions.
We acknowledge support from the Office of Nuclear Physics in the
Office of Science of the Department of Energy,
the National Science Foundation,
Abilene Christian University Research Council,
Research Foundation of SUNY, and
Dean of the College of Arts and Sciences, Vanderbilt University
(U.S.A),
Ministry of Education, Culture, Sports, Science, and Technology
and the Japan Society for the Promotion of Science (Japan),
Conselho Nacional de Desenvolvimento Cient\'{\i}fico e
Tecnol{\'o}gico and Funda\c c{\~a}o de Amparo {\`a} Pesquisa do
Estado de S{\~a}o Paulo (Brazil),
Natural Science Foundation of China (People's Republic of China),
Croatian Science Foundation and
Ministry of Science and Education (Croatia),
Ministry of Education, Youth and Sports (Czech Republic),
Centre National de la Recherche Scientifique, Commissariat
{\`a} l'{\'E}nergie Atomique, and Institut National de Physique
Nucl{\'e}aire et de Physique des Particules (France),
Bundesministerium f\"ur Bildung und Forschung, Deutscher Akademischer
Austausch Dienst, and Alexander von Humboldt Stiftung (Germany),
J. Bolyai Research Scholarship, EFOP, the New National Excellence
Program ({\'U}NKP), NKFIH, and OTKA (Hungary),
Department of Atomic Energy and Department of Science and Technology
(India),
Israel Science Foundation (Israel),
Basic Science Research and SRC(CENuM) Programs through NRF
funded by the Ministry of Education and the Ministry of
Science and ICT (Korea).
Physics Department, Lahore University of Management Sciences (Pakistan),
Ministry of Education and Science, Russian Academy of Sciences,
Federal Agency of Atomic Energy (Russia),
VR and Wallenberg Foundation (Sweden),
the U.S. Civilian Research and Development Foundation for the
Independent States of the Former Soviet Union,
the Hungarian American Enterprise Scholarship Fund,
the US-Hungarian Fulbright Foundation,
and the US-Israel Binational Science Foundation.

\end{acknowledgments}




%
 
\end{document}